\documentclass[12pt]{article}
\usepackage{graphicx}
\begin{document}
\newcommand{\gsim}{\raisebox{-0.8ex}{\mbox{$\stackrel{\textstyle>}{\sim}$}}}
\newcommand{\lsim}{\raisebox{-0.8ex}{\mbox{$\stackrel{\textstyle<}{\sim}$}}}
\newcommand{\re}[1]{(\ref{#1})}
\newcommand{\dl}{^{\rm dl}}
\newcommand{\dll}{_{\rm dl}}
\newcommand{\msp}{M_{\rm sph}}
\newcommand{\esp}{E_{\rm sph}}
\newcommand{\rsp}{R_{\rm sph}}
\newcommand{\dsp}{D_{\rm sph}}
\newcommand{\msk}{M_{\rm sk}}
\newcommand{\rsk}{R_{\rm sk}}
\newcommand{\esk}{E_{\rm sk}}
\newcommand{\dsk}{D_{\rm sk}}
\renewcommand{\max}{M_{\rm axial}}
\newcommand{\mm}{M_{\rm mon}}
\newcommand{\ma}{m_{\rm A}}
\newcommand{\mps}{m_\Psi}
\newcommand{\dm}{D_{\rm mon}}
\newcommand{\mv}{m_{\rm V}}
\newcommand{\mh}{m_{\rm H}}
\newcommand{\agut}{\alpha_{\rm GUT}}
\newcommand{\eins}{1\hspace{-0.56ex}{\rm I}}

\newcommand{\1}{\cite{brtz}}
\newcommand{\fax}{f_\infty}
\newcommand{\be}{\begin{equation}}
\newcommand{\ee}{\end{equation}}
\newcommand{\bea}{\begin{eqnarray}}
\newcommand{\eea}{\end{eqnarray}}
\newcommand{\const}{\mathop{\rm const}\nolimits}

\title{ \bf An extended
model for monopole catalysis of nucleon decay}
\author{
{\large Y.~Brihaye}$^{\diamond}$, 
{\large D.~Yu.~Grigoriev}$^{\dagger,\star}$,
{\large V.~A.~Rubakov}$^{\dagger}$ and \\
  {\large D.~H.~Tchrakian}$^{\star}$\\ \\
$^{\diamond}${\small Physique-Math\'ematique, Universit\'e de 
Mons-Hainaut, Mons, Belgium}\\ \\
$^{\dagger}${\small Institute for Nuclear Research of the
Russian Academy of Sciences,} \\
{\small 60th October Anniversary Prospect, 7a, 117 312, Moscow,
Russia}\\ \\
$^{\star}${\small Mathematical Physics, National University of Ireland
Maynooth (NUIM),} \\
{\small Maynooth, Co. Kildare, Ireland}}
\date{}

\maketitle

\begin{abstract}
A new model for monopole catalysis of nucleon decay is proposed.
Unlike in the earlier one,  the only light fields
in this model are the photon and Skyrme (pion) field.
The model admits the 't~Hooft--Polyakov monopole and Skyrmion as 
classical solutions, while baryon number non-conservation occurs
through an anomaly involving an intermediate mass axial vector field
resembling $W$- and $Z$-bosons. By considering spherically
symmetric monopole-Skyrmion configurations, we find that the Skyrmion
looses essentially all its mass when interacting with the monopole,
which is phenomenologically identical to the monopole catalysis of
nucleon decay. Short-distance monopole-Skyrmion physics in this model
is interesting too, as there exist almost degenerate metastable
monopole-Skyrmion bound states separated by substantial energy
barriers. Yet the heights of the latter are much smaller than the
physical nucleon mass, so the complete disappearance
 of the normal undeformed Skyrmion
remains perfectly possible in Skyrmion-monopole scattering.
\end{abstract}

\section{Introduction}

Models admitting both 't~Hooft--Polyakov monopole~\cite{mon-t,mon-p} 
and Skyrmion~\cite{skyrme} as static solutions to field equations are
of interest from various points of view. One aspect is the possibility
to describe the monopole catalysis of nucleon decay~\cite{R,C}
at purely classical level~\cite{CW,brtz}. In Grand Unified Theories, 
dynamics of the monopole catalysis involves several widely separated
length scales, the smallest of which is the size of the monopole core,
roughly of the order of $1/v$, where $v$ is the GUT vacuum expectation
value (vev), $v \sim 10^{16}$~GeV, while the largest one is the
hadronic scale of the order of $1/F_{\pi}$ where $F_\pi \sim
100$~MeV. One of the sources of baryon number non-conservation, which
may be relevant to the monopole catalysis, is the electroweak 
anomaly~\cite{anom}, with the corresponding length scale roughly of
the order of $1/\zeta$ where $\zeta \sim 100$~GeV is the Standard
Model vev. The existence of so different scales, as well as the
non-perturbative nature of dynamics, make any analysis of the monopole
catalysis difficult.

A simple model aimed to describe the catalyzed decay of a nucleon,
represented by the Skyrmion, in interactions with the 't~Hooft--Polyakov
monopole was suggested in Ref.~\1. That model, however, featured in
its spectrum an axial vector boson with mass scale set by $F_\pi$
rather than by the electroweak scale. This unrealistic property is
particularly unpleasant in view of the fact that baryon number 
non-conservation in that model occurs via anomalous interactions of
the Skyrme field to the axial vector boson, so one might worry that
the Skyrmion decay in the presence of the monopole is unsuppressed
precisely because the axial vector field is so light.

In the present work the simplified model of nucleon decay \1 is
refined by addition of a new axial symmetry-breaking Higgs field in
the same group representation as the Skyrme field itself. The vev 
$\zeta$ of
the extra field introduces a new energy scale into the model which 
 naturally resembles
the electroweak scale, and hence the new
massive axial bosons resemble the $W$- and $Z$-bosons. 
Thus, the only light fields in the extended model are the Skyrme
(pion) field and the electromagnetic field. In this sense the model is
completely realistic.

In this paper we study static, spherically symmetric field
configurations in the extended model. Our main findings are
two-fold. First, in complete analogy to Ref.~\1 we observe that on top
of the monopole, there are no static solutions with baryon number of
order one, whose mass and/or size are comparable to the mass and/or
size of a nucleon (``bare'' Skyrmion). This is directly relevant to
the monopole catalysis: whatever happens to nucleons/quarks near the
monopole core, the nucleon gets squeezed into a small region near the
monopole, and its mass is irradiated away in the form of pions.
This is phenomenologically identical to the catalyzed nucleon decay.

The second set of observations concerns the properties of our system
at short distances, i.e., near the monopole core. Of course, the
description of the strong interaction physics in terms of the Skyrme
model is not valid at  short distances, yet our results may be of
interest for understanding the Skyrmion-monopole system per se. We
find, in sharp contrast to Ref.~\1, that
the unimpeded decay of the static Skyrmion-like state is no
longer possible; instead we find a sequence of Skyrmion-monopole bound
states separated by energy barriers, the picture 
similar to the familiar periodic
structure of topological vacua
in gauge  theories. The energy minima correspond to integer
baryon charge and their energies (referenced from the mass of the
``bare'' monopole) are of order $F_\pi^2/(gv)$, up to logarithm, where
$g$ is the gauge coupling constant.  The height 
of the energy barrier at half-integer
baryon charge is defined by the solution involving 
the extra Higgs field,
analogous to the sphaleron solution~\cite{KM}.
At $\zeta \gg F_\pi$, where $\zeta$ is the intermediate energy scale
(vev of the extra Higgs field), the barrier height is of order
$\zeta^2/(gv)$. Even though the barrier height is large compared to
the masses of metastable bound states, all energies involved are tiny
compared to the physical nucleon mass, provided the realistic
hierarchy
\[
  F_\pi \ll \zeta \ll v
\]
holds. This indeed demonstrates that the nucleon looses
almost all its energy when interacting with the monopole.

For the sake of completeness we consider also
the  unphysical case
of large $\zeta/v$. In that case
the extra Higgs field modifies the
Chern-Simons number in a nontrivial way, resulting in the multiple
valued dependence of the baryon number
$B$ on the asymptotic value of the Skyrme field
$\fax$. The dependence of energy $E$
on $B$ is still uniquely defined, with sharp minima at integer $B$,
but it is no longer possible to identify a single well-defined barrier
separating these minima.

It is worth noting that the static results obtained both in the
simplified model~\1 and its extended version described below are not
directly applicable to the kinetics of real Skyrmion decay, once the
latter is not even a solution in the static models. The object
stabilised by the introduction of the intermediate 
scale $\zeta$ is a
static Skyrmion-like solution localized on the monopole core with
energy negligible compared to that of real Skyrmion. In other
words, the Skyrmion decay is a non-stationary process; we hope to
study its dynamics
in a future publication.

\section{The Model}

\subsection{Field content}

As explained in Ref.~\1, our model is designed to leave an unbroken
(electromagnetic) $U(1)$. It consists of the Skyrme field $U$ coupled
to two gauge fields:
\[
    SU(2)_L :  \quad A_{\mu}
\]
\[
     SU(2)_R : \quad B_{\mu}
\]
In addition to the two Georgi-Glashow Higgs fields $\Phi_A$ and
$\Phi_B$ which break the $SU(2)_L$ and $SU(2)_R$ symmetries 
down to $U(1)_L$ and $U(1)_R$,
respectively, at the ``GUT'' scale $v$,
we add a third Higgs field $\Psi$ which breaks the axial subgroup of
the remaining $U(1)_L \times U(1)_R$. 
In the original model \1 the axial $U(1)$ had been broken by
the Skyrme field itself, with the corresponding vector boson mass of the
order of $gF_{\pi}$. Here the scale of axial symmetry breaking is
fixed by the vacuum expectation value of the $\Psi$-field, 
$\zeta$, which we assume to be 
an intermediate scale between $v$ and $F_\pi$.
To break  the axial $U(1)$ only, $\Psi$-field must be
in the same group representation as the Skyrme field. The latter
requirement fixes the scalar field content of the updated model
completely:
\begin{eqnarray}
  U \in SU(2)\; , \;\;\; U,\Psi&:& \;\;({\bf 2}, {\bf \bar{2}}) \quad  \quad  \quad \nonumber \\
    \Phi_A &:& \;\; ({\bf 3}, {\bf 1}) \nonumber \\
    \Phi_B &:& \;\; ({\bf 1}, {\bf 3})
\end{eqnarray}
As in Ref.~\1, we take the gauge couplings, 
triplet Higgs self-couplings and their vacuum 
expectation values to be the same in the left ($L$)
and right ($R$) sectors,
\[
    g_A = g_B \equiv g \; , \;\;\;
    \lambda_A = \lambda_B \equiv \lambda \; , \;\;\;
    v_A = v_B \equiv v
\]
The action for this model is
\begin{eqnarray}
     S &=& \int~d^4x~\left[ -\frac{1}{2g^2} \mbox{Tr}(F_{\mu\nu}^2)
 -\frac{1}{2g^2} \mbox{Tr}(G_{\mu\nu}^2)\right] \nonumber \\
  &-& \int~d^4x~\left[
\frac{1}{2} \mbox{Tr}(D_{\mu}\Phi_A)^2 + 
\frac{1}{2} \mbox{Tr}(D_{\mu}\Phi_B)^2
\right]
\nonumber \\
&-&  \int~d^4x~\left[
\frac{\lambda}{4} 
\left(\frac{1}{2} \mbox{Tr}(\Phi_A^2) + v^2\right)^2
+  \frac{{\lambda}}{4} 
\left(\frac{1}{2}\mbox{Tr}(\Phi_B^2) + v^2\right)^2
\right] \nonumber \\
&+& \int~d^4x~\left[\frac{1}{2}\mbox{Tr}(D_{\mu}\Psi^\dagger
D_{\mu}\Psi) 
+
\frac{\tau}{4}\left(\frac{1}{2}\mbox{Tr}(\Psi^\dagger\Psi)-\zeta^2\right)^2
\right]\nonumber\\
&+& \int~d^4x~\left[  
- \frac{F_{\pi}^2}{16} \mbox{Tr}(U^{\dagger} D_{\mu} U)^2
 + 
\frac{1}{32e^2}\mbox{Tr}([U^{\dagger}D_{\mu}U,U^{\dagger}D_{\nu}U]^2)
\right] \nonumber \\
&+& \Gamma_{WZW}
\label{action}
\end{eqnarray}
Here $F_{\mu\nu}$ and $G_{\mu\nu}$ are the field strengths
of $A_{\mu}$ and $B_{\mu}$, the fields $\Phi_A$ and $\Phi_B$
are $2 \times 2$ matrices from the algebras of $SU(2)_L$ and 
$SU(2)_R$, respectively,
\begin{eqnarray}
    D_{\mu}U &=& \partial_{\mu} U + A_{\mu}U - UB_{\mu}
      \nonumber \\
    D_{\mu} \Phi_A &=& 
     \partial_{\mu}\Phi_A + [A_{\mu}, \Phi_A]
\nonumber \\
D_{\mu} \Phi_B &=& 
     \partial_{\mu}\Phi_B + [B_{\mu}, \Phi_B]
\end{eqnarray}
and $F_{\pi}$ and $e$ are the pion decay constant and the Skyrme
constant. Generally speaking, the new Higgs field $\Psi$
is an arbitrary $2 \times 2$ complex matrix with covariant derivative
identical to that  of the Skyrme field, $$D_{\mu}\Psi = \partial_{\mu}
\Psi + A_{\mu}\Psi - \Psi B_{\mu}$$  
As in Ref.~\1, in what follows we can completely ignore the 
Wess--Zumino--Witten term
$\Gamma_{WZW}$
in \re{action}.

\subsection{Masses and energies}

For the physical hierarchy  $v \gg \zeta \gg
F_{\pi}$, the physical spectrum of the system decouples into a set of
well-defined objects with masses of relevant scales. The heaviest mass
scale is controlled by $v$ -- the vacuum expectation value of the ``GUT'' 
Higgs
fields $\Phi_A$ and $\Phi_B$. Here we have ``GUT'' 
Higgs and vector boson
masses $$ \mv=2gv$$ $$\mh=\sqrt{\lambda}v $$ and monopole mass
\begin{equation}
   \mm= 4\pi\dm \frac{v}{g}
\label{12**}
\end{equation}
where $\dm$ at ${\lambda}/g^2 = 0.5$ is around 2.4 . 

The next mass scale is related to $\zeta$ -- the vacuum expectation
value of the extra Higgs field $\Psi$. In the proper vacuum state (see
below) the extra field adds mass to the axial $U(1)$ gauge boson only:
\be
\ma = \left(g^2\zeta^2 + {g^2F_\pi^2\over 8}\right)^{1/2}
\ee
while the mass of the extra Higgs boson is
\be
\mps=\sqrt{\tau}\zeta
\ee
The mass of corresponding non-perturbative object would have been of
the
order of 
\[
\msp\sim\frac{\zeta}{g}
\]
 Once the axial gauge field is meant to mimic $W$- and $Z$-bosons,
the non-perturbative object should be analogous to
the sphaleron. As we will show below, this object indeed exists and
defines the barrier height between local minima of static energy
with integer
baryon number. However, as will be shown below, in the presence of
the monopole the corresponding unstable solution 
becomes much lighter than $\msp$:
\be
 \esp \sim {\msp^2\over\mm} \sim \frac{\zeta^2}{gv}
\ee
This is similar to what happens to the Skyrmion in our model (see also \1).
In the absence of pion mass potential term, the normal (``bare'') 
Skyrmion \cite{skyrme} would have mass on the $F_\pi$ scale,
$$\msk=\frac{4\pi}{e}\dsk F_\pi, \quad \dsk\sim 2.9$$ 
and radius 
$$\rsk\sim \frac{1}{eF_\pi}$$ 
We will see that in the monopole background, 
the normal Skyrmion is no longer
stable, which by itself points to the possibility of nucleon decay
in the model at hand. The new ``thin'' Skyrmion (the static solution
with unit baryon charge) is  considerably smaller and lighter:
\be
\rsk'\sim \frac{1}{gv}
\ee
\be
\esk\sim{\msk^2\over\mm} \sim \frac{F_\pi^2}{gv}
\ee

\subsection{Spherically symmetric Ansatz}
As in Ref.~\1, we can simplify the Ansatz because the 
action (\ref{action}) is invariant under the spatial reflection,
$x^0 \to x^0$, ${\bf x} \to - {\bf x}$ supplemented by the 
interchange of the left and right $SU(2)$, i.e.,
\begin{eqnarray}
   A_0 &\leftrightarrow& B_0\; , \;\;\; A_i \to -B_i\; ,
  \;\;\; B_i \to - A_i \nonumber \\
   \Phi_A &\to& - \Phi_B\; , \;\;\;  \Phi_B \to - \Phi_A\;\;\; ,
   U \to U^{\dagger}\; , 
\;\;\; \Psi \to \Psi^{\dagger}
\label{discretesymmetry}
\end{eqnarray}
(all numerical solutions found are consistent with this symmetry. For
static fields with $A_0 = B_0 = 0$, the most general spherically
symmetric Ansatz consistent with (\ref{discretesymmetry}) is
\begin{eqnarray}
A_i &=& -{i\over 2}
\left[ \left( \frac{a_1 (r)-1}{r}\right)\varepsilon_{ijk}\sigma_j \hat x_k
+\left( \frac{a_2 (r)}{r} \right) 
(\sigma_i -\hat x_i \hat x \cdot \vec \sigma )
+\left( \frac{a_3 (r)}{r} \right) 
\hat x_i \hat x \cdot \vec \sigma \right] \nonumber \\
B_i &=& -{i\over 2}
\left[ \left( \frac{a_1 (r)-1}{r}\right)\varepsilon_{ijk}\sigma_j \hat x_k
-\left( \frac{a_2 (r)}{r} \right) 
(\sigma_i -\hat x_i \hat x \cdot \vec \sigma )
-\left( \frac{a_3 (r)}{r} \right) 
\hat x_i \hat x \cdot \vec \sigma \right]\:
\nonumber \\
\Phi_A &=& \Phi_B = i v h(r) \hat x \cdot \vec \sigma
\nonumber \\
U&=&\eins\cos f(r) +i\hat x \cdot \vec \sigma \sin f(r)\;\;\; ,\;\;\;
\Psi=\zeta[ih_1(r)\hat x.\sigma+h_2(r)\eins]
\label{6m**}
\end{eqnarray}
where $\hat x$ is the unit radius-vector.

It is worth noting that in the spherically-symmetric Ansatz (\ref{6m**})
the field $\Psi$ automatically breaks the axial subgroup of $U(1)_L
\times U(1)_R$ and preserves vectorial subgroup, so the photon remains
massless. In less symmetric cases
this happens only for a subset of all possible $\Psi$-field vacua, so
it would be necessary to include
additional terms into action to single out the correct 
symmetry breaking pattern yielding a massless photon.

With the Ansatz (\ref{6m**}), regularity of the Skyrme field
at the origin requires that $f(0)$ is an integer 
multiple of $\pi$;
without loss of generality we set
\[
    f(0) = \pi
\]
Other conditions, ensuring that the fields are regular at
the origin, are
\[
  a_1 (0) = 1\;, \;\;\;  a_2 (0) = a_3 (0) =0 \;, \;\;\; 
  h(0) = h_1(0) = h_2'(0) = 0
\]

Using the dimensionless coordinate
\[
   \rho = gvr
\]
and neglecting the WZW
term in Eq.~(\ref{action}), the static energy 
functional can be written as follows:
\[
  H =  \frac{4\pi v}{g}  \int~d\rho~\cal{E}(\rho)
\]
\begin{eqnarray}
  \cal{E} &=& \left[ \left( a_1'+\frac{a_2 a_3}{\rho}\right)^2 +\left(
a_2'-\frac{a_1 a_3}{\rho}\right)^2 +\frac{1}{2\rho^2}( a_1^2 +a_2^2 -
1)^2 \right] \nonumber \\ 
&+& 2 \left[\rho^2 (h')^2 +2 (a_1^2 +a_2^2) h^2
+\frac{{\lambda}}{4g^2} \rho^2(h^2-1)^2\right]\nonumber \\
&+& \left({\zeta\over v}\right)^2 \left[\rho^2 
\left(h_1'-\frac{a_3}{\rho}h_2\right)^2 +
\rho^2 \left(h_2'+\frac{a_3}{\rho}h_1\right)^2  \right. \nonumber \\
& & \left. \quad \quad {}+ 2 (a_1h_1-a_2h_2)^2
+\frac{{\tau}}{4g^2} \left({\zeta\over v}\right)^2 \rho^2
(h_1^2+h_2^2-1)^2\right]\nonumber \\ 
&+& \kappa_1 (X^2 + 2Y^2) + \frac{4\kappa_2}{\rho^2} Y^2 (2X^2 + Y^2)
\label{energyrad}
\end{eqnarray}
where the prime denotes the derivative with respect to $\rho$, 
\begin{eqnarray}
    X &=& \rho f' - a_3
   \nonumber \\
    Y &=& a_1 \sin f - a_2 \cos f\;\;\; ,
\label{defXY}
\end{eqnarray}
and 
\begin{eqnarray}
  \kappa_1 &=& \frac{F_{\pi}^2}{8 v^2}
\nonumber \\
  \kappa_2 &=& \frac{g^2}{64e^2}
\label{kappas}
\end{eqnarray}

The radial energy functional (\ref{energyrad}) retains a residual gauge
invariance, which is a remnant of the axial $U(1)$:
\begin{eqnarray}
h_2+ih_1 &\to& (h_2+ih_1) e^{i\alpha(\rho)} \nonumber \\
a_1+ia_2 &\to& (a_1+ia_2) e^{i\alpha(\rho)} \nonumber \\
    a_3 &\to& a_3 + \rho \partial_\rho \alpha(\rho) \nonumber \\
    f &\to& f + \alpha(\rho)
\label{restgauge}
\end{eqnarray}
Note that $h_2+ih_1$ rotates identically to $e^{if}$, once the 
fields $\Psi$
and $U$  are in the same group representation.

Making use 
of~(\ref{energyrad}) and (\ref{defXY}), one finds the boundary
conditions at $\rho\rightarrow\infty$ from the requirement of finite
energy. In the absence of the monopole we would have 
$a_1=1,\ a_2=0$, and hence 
$\fax=\pi N$. The latter condition would 
prevent the Skyrmion
from decaying, thus making the baryon number $B$ topologically stable.

The presence of the monopole requires that the ``GUT'' Higgs fields be
topologically nontrivial at infinity:
$$ h(\infty)=1$$ 
which immediately leads to $a_1(\infty)=a_2(\infty)=0$, so that now the
Skyrme field at infinity $\fax$ can take any value, allowing for the
Skyrmion decay, while
$a_3(\infty)=(\rho f')_{\rho\rightarrow\infty}$.

The boundary conditions for $\Psi(\infty)$ require more careful
consideration. The only condition following from (\ref{energyrad})
is $$\left(h_1^2+h_2^2\right)_{\rho\rightarrow\infty}=1$$ or
\be
\left(h_2+ih_1\right)_{\rho\rightarrow\infty}=e^{i\beta}
\label{hax}
\ee
where $\beta=0$ in the vacuum. 
One can see immediately from (\ref{restgauge}) that the
difference $\fax-\beta$ is gauge invariant. Let us show that this
quantity is also time independent  in the dynamical version of the
model. This is clear (see also Ref.~\cite{brtz}) in the $a_0=b_0=0$ gauge,
where the canonical momenta are proportional to time derivatives of
the fields. The kinetic energy of the Higgs and Skyrme fields contains
the terms $$ \int r^2dr\,\dot{h}^2 , \quad \int
r^2dr\,\left(\dot{h}_1^2+\dot{h}_2^2\right) \quad\mbox{and} \quad \int
r^2dr\,\dot{f}^2$$ therefore during any dynamical evolution the time
derivatives of fields $h$, $h_1$, $h_2$ and $f$ must vanish at
infinity, so that their asymptotic values are time
independent.\footnote{This is not the case for the field $a_3$ which
enters the kinetic energy through $F_{0i}\propto \dot{a}_3/r$. } Once
$\fax-\beta$ is gauge invariant, it is time independent in any gauge,
including the $a_3=0$ gauge extensively used below.

Introduction of the new field $\Psi$ does not alter the expression for
baryon number $B$. In the spherically symmetric case one has
\[
   B= \int_0^{\infty}~d\rho~ b(\rho)
\]
\begin{eqnarray}
 b(\rho) &=& \frac{1}{\pi}  \frac{d}{d\rho}
\left[ -f + \frac{1}{2} (a_1^2 - a^2_2) \sin 2f 
- a_1 a_2 \cos 2f \right]
\nonumber \\
&+& \frac{1}{\pi} \left[(a_1a_2' - a_1' a_2) 
- \frac{a_3}{\rho} (a_1^2 + a_2^2 -1) \right]
\label{bnumber}
\end{eqnarray}
which in the gauge $a_3=0$ takes the form
\begin{equation}
   B = \frac{\pi - f_{\infty}}{\pi} + 
\frac{1}{\pi} \int_0^{\infty}~d\rho~(a_1 a_2^{\prime} -
a_1^{\prime} a_2)
\label{bv*}
\end{equation}
Therefore, in the $a_3=0$ gauge the baryon number $B$ is parameterized by
the free parameter $\fax$, with $\fax=\pi$ at $B=0$. By finding a set of static solutions
corresponding to different values of $\fax$, we build a possible
trajectory for quasistationary Skyrmion decay (for more detailed
discussion see \cite{brtz}). 
Recalling that $\beta-\fax$ is constant during the decay,
we finally obtain the boundary condition for $\Psi$:
\be
\beta=\fax-\pi
\label{beta} 
\ee
As will be seen below, the condition (\ref{beta}) plays a crucial
role. Without it, the $\Psi$-field at any $B$
remains close to its vacuum configuration and effectively decouples
from the rest of the system whose properties 
would remain similar to those
described in Ref.~\cite{brtz}.

Finally, let us write the energy functional in $a_3=0$ gauge:
\begin{eqnarray}
  \tilde{H} & \equiv & H(a_3=0) 
=  \frac{4\pi v}{g}  \int~d\rho~\tilde{\cal{E}}(\rho) \nonumber \\
  \tilde{\cal{E}} &=& 
\left[ ( a_1')^2
+( a_2')^2 +\frac{1}{2\rho^2}( a_1^2 
+a_2^2 - 1)^2
\right] \nonumber \\
&+&
2 \left[\rho^2 (h')^2 +2 (a_1^2 +a_2^2) h^2
+\frac{{\lambda}}{4g^2} \rho^2(h^2-1)^2\right]\nonumber \\
&+& \left({\zeta\over v}\right)^2 
\left[\rho^2 \left( (h_1')^2 + (h_2')^2 \right) +2 (a_1h_1-a_2h_2)^2
  \right.\label{energyrad1} \\
& & \quad \quad \quad \quad \quad \quad \left. + \frac{{\tau}}{4g^2} \left({\zeta\over v}\right)^2 \rho^2
(h_1^2+h_2^2-1)^2\right]\nonumber \\
&+& \kappa_1 (\rho^2 (f')^2 + 2Y^2) + \frac{4\kappa_2}{\rho^2} Y^2 (2\rho^2 (f')^2 + Y^2)\nonumber
\end{eqnarray}  
with $Y$ defined by (\ref{defXY}). Eq.~\re{energyrad1} is the energy
density functional that will be used in the subsequent study.

\section{Static solutions}

\subsection{Overall picture}

The energy (15), referenced from the
monopole mass, is shown in Fig.~\ref{added} as a function of the baryon number.
This plot is obtained numerically in a way outlined at the end of
the previous section, and  has always the same shape in the
physical range of parameters, $F_\pi \ll \zeta \ll v$. One observes
that there indeed exist local minima at integer $B$ separated by
barriers centered at half-integer baryon number. The corresponding
values of the energy are much lower than the mass of the bare Skyrmion, which
in units of Fig.~\ref{added} is equal to 1.

It is
interesting to note that the increase in $\zeta/v$ gives rise to a
considerable contribution to the Chern-Simons integral in (\ref{bv*}),
so that the curve $B(f_\infty)$ is no longer monotonic. This results in
the bifurcations exhibited on the $E(f_\infty)$ plot of Fig.~\ref{fig2}a.
However, even in the regime $\zeta \gg v$, the energy remains a
single valued function of the  baryon charge $B$ (see
Fig.~\ref{fig2}b).

Let us come back to the physical range of parameters.
Like the
usual sphaleron barriers separating topologically inequivalent vacua
in electroweak theory \cite{KM,Rubakov:1996vz}, the barriers 
separating states with integer baryon number are of
topological origin. The vacuum configuration of the field $\Psi$ is
$h_1=0, h_2=1$, but at noninteger $B$ the boundary conditions for the
field $\Psi$ lead to its departure from this
vacuum. As mentioned above,
the gauge invariant difference $\fax-\beta$
is fixed and is equal to $\pi$ along physical evolution
paths. In the $a_3=0$ gauge, integer $B$
corresponds to $\fax$ being a multiple of $\pi$ (this is the case even
for large $\zeta/v$, when the Chern-Simons term in Eq.~(\ref{bv*}) is
generally not negligible). Therefore, at integer $B$ the boundary
conditions for the $h_1$ component of the $\Psi$-field
$$ h_1(0)=h_1(\infty)=0 $$ are compatible with the vacuum configuration,
and the contribution of $\Psi$ to the total energy is small. 

This is no longer the case at non-integer $B$, when $h_1(\infty)\neq
0,$ and $h_1(0)$ is still 0. In other words $h_1(\infty)\neq 0$ leads to
a hedgehog configuration of the $\Psi$-field at infinity, giving rise to a
topological defect which in the  spherically symmetric case is
located at the origin. The contribution of this object to the energy
and the
$E(B)$ curve giving the  shape of the barrier
can be evaluated in the spectator approximation, defined below.

\subsection{Spectator approximation}

\subsubsection{Energy of the Skyrme field}

For the physical range of the parameters in the Hamiltonian
(\ref{energyrad1}), there is a hierarchy of ``coupling constants'' 
which
reflects the presence of three physical scales in the problem,
\[
 \frac{F_\pi^2}{8v^2} \equiv 
\kappa_1 \ll {\zeta^2\over v^2} \ll \kappa_2 \sim 1$$,
which means that the monopole is so heavy that the effect of the other
fields on it is negligible. In Ref.~\cite{brtz} this was called the
spectator regime. It sets in at $\kappa_1/\kappa_2\;
\lsim\; 0.01$, see Fig.~\ref{fig1}.  

An important property of the spectator regime is that in the $a_3=0$
gauge,
the low-energy
fields $f$ and $\Psi$ can be evaluated in the background field of the
monopole, so they  decouple
and depend only on their own ``coupling constants''
$\kappa_1/\kappa_2\propto{(F_{\pi}/v)^2}$ and $(\zeta/v)^2$
respectively. As shown in
Ref.~\cite{brtz}, for $\rho\; \gsim\; \rho_c \sim 1$
the solution $f(\rho)$ has simple form,
\begin{equation}
   f(\rho) = f_{\infty} + (\pi-f_{\infty}) \frac{\rho_c}{\rho}
\label{a**}
\end{equation}
where
\begin{equation}
   \rho_c \sim \frac{1}{4} 
\left| \ln \left(\frac{\kappa_1}{\kappa_2}\right)\right|
\label{13m*}
\end{equation}
Substituting (\ref{a**}),~(\ref{13m*}) into the Hamiltonian
(\ref{energyrad1}), one obtains an estimate for 
the contribution of the Skyrme field to the energy
in the
presence of the monopole: 
\be
\esk \sim {{4\pi v}\over g}\kappa_1\rho_c(\pi-f_\infty)^2
= \frac{\pi^3 F_\pi^2}{8vg}
\left|\ln{{\kappa_1}\over{\kappa_2}}\right|
B^2 
\label{eskyrm}
\ee
where we made use of the relation
\be
B=1-{\fax\over\pi}=-{\beta\over\pi}
\label{bbb}
\ee
valid in the spectator approximation.

\subsubsection{Energy of extra Higgs field and the sphaleron}

A similar estimate can be made for the contribution of the 
$\Psi$-field. 
Supposing that $h_2=\const=\cos\beta$ over all the volume, one
obtains 
the following general
equation for $h_1$,
\be
(\rho^2h_1')'-2a_1^2h_1 = {\tau\over{2g^2}}\left({\zeta\over
v}\right)^2\rho^2\left(h_1^2-\sin^2\beta \right)h_1
\label{generaleq}
\ee
 Outside the monopole core, where
$a_1=0$, one  linearizes  near $h_1=\sin \beta$, and finds
\[
  h_1 (\rho)-\sin\beta
= \frac{\mbox{const}}{\rho} 
\exp\left(\pm \frac{\sqrt{\tau}\zeta}{gv} \sin \beta\cdot \rho \right)
\]
The minus sign gives a solution that remains finite at $\rho
\rightarrow \infty$. 
Taking into account that
$\zeta/v \ll 1$, one obtains 
\be
h_1 = \sin \beta + {\const \over \rho}
\label{asph1}
\ee
Now let us find $h_1$ near the monopole centre.
For small $\rho$, Eq.~(\ref{generaleq}) reduces to 
\[
(\rho^2h_1')'=2a_1^2h_1
\]
Given that $a_1(0) = 1$, the solution near the centre
is
\be
h_1 = \const\cdot\rho
\label{asph2}
\ee
To obtain the approximate solution everywhere, 
we match the two expressions for $h_1$,
namely (\ref{asph1}) and (\ref{asph2}), as well as their derivatives, at
$\rho=\rho_0 \sim 1$ and find
\be
\begin{array}{lll}
\displaystyle
h_1 ={\rho\over{2\rho_0}} \sin \beta  \hfill        & & \rho<\rho_0
\\ \displaystyle
h_1 =\left(1-{\rho_0\over{2\rho}}\right) \sin \beta
        & & \rho>\rho_0 
\end{array}
\label{sph}
\ee
The best agreement with numerics (see Fig.~\ref{fig4}) is obtained at
$\rho_0\sim 0.76$. Substituting (\ref{sph}) into the Hamiltonian
(\ref{energyrad1}), one finds 
for the contribution of the $\Psi$-field to the energy
\be
E_{\Psi}=\esp  \sin^2 \beta
\label{periodic}
\ee
where
\be
\esp\sim 4\pi{{\zeta^2}\over{gv}} \, {{\rho_0}\over 2}
\label{esph}
\ee which is about 50\% higher than the actual numerical value, see
Fig.~\ref{fig5}. It is worth noting that, similarly to the Skyrmion, the
solution is stabilized against shrinking to zero size
only by the monopole core. For a
pointlike monopole $\rho_0\rightarrow 0$ and $\esp\rightarrow 0$.

Taking into account the relation (\ref{bbb}),
valid in the spectator approximation,
one can rewrite
Eq.~(\ref{periodic}) as
\be
E_{\Psi} (B)=\esp \sin^2 \pi B
\label{periodic1}
\ee
which is indeed similar \cite{Rubakov:1996vz} to the periodic
vacuum structure of the standard electroweak theory.
Local maxima of $E_\Psi (B)$ at half-integer $B$ correspond to
sphaleron-like solutions in the presence of the monopole.

\subsubsection{Total energy}

In the spectator approximation, the curve
$\Delta E (B)=E(B)-\mm$ is the sum of the two contributions
(\ref{eskyrm}) and
(\ref{periodic1}),
\be
 \Delta E (B) = C_1 \frac{F_\pi^2}{gv}\cdot B^2 + 
C_2 \frac{\zeta^2}{gv} \sin^2 \pi B
\label{deltae}
\ee
where aproximate expressions for
$C_1$ and $C_2$ are determined by (\ref{eskyrm}) and
(\ref{esph}), respectively,
and their values are
of order one (up to logarithm).
 Comparison with numerical data is shown in
Figs.~\ref{fig6},\ref{fig7}. Even though both terms are suppressed by
the ``GUT'' vev $v$, the sphaleron remains much heavier than the 
Skyrmion bound state, since $\zeta\gg F_{\pi}$. 
In other words, the energy gain
from the decay of the Skyrmion bound state
 is of the order of $\esk\sim
F_{\pi}^2/(gv)$, which is much less than the barrier height $\esp\sim
\zeta^2/(gv)$. Therefore, the introduction of intermediate energy scale
prohibits the decay of the Skyrmion-like solution along the minimal
energy path.

\section{Conclusions}
In the model employed in \1 the axial vector mass was unphysically light.
Here we have rectified this by the
introduction of an extra Higgs field with vev $\zeta$ which is meant
to
mimick the
electroweak energy scale. A continuous set of static solutions 
with varying baryon number
is found numerically. 
The extra Higgs field gives rise to energy barriers along this set.

In the physical range of the parameters of the model we have a
complete understanding of its static properties, with good agreement
between analytic estimates and numerical data. 
The monopole is so heavy that its deformations due to interaction with
the rest of the system are negligible. In a convenient gauge $a_3=0$,
the Skyrme field and extra Higgs field decouple\footnote{Except that they are related through the boundary condition (\ref{beta}) at spatial infinity.},
their  shapes are defined by  interaction with
the background monopole. Both the Skyrmion and sphaleron 
are strongly deformed by the
monopole: their asymptotic behaviour emerges immediately outside the
monopole core, and they collapse to the origin in the limit of pointlike
monopole. The energies of these solutions are suppressed by the
largest vev.

\begin{figure}[htb]
\centering \includegraphics[height=3.5in]{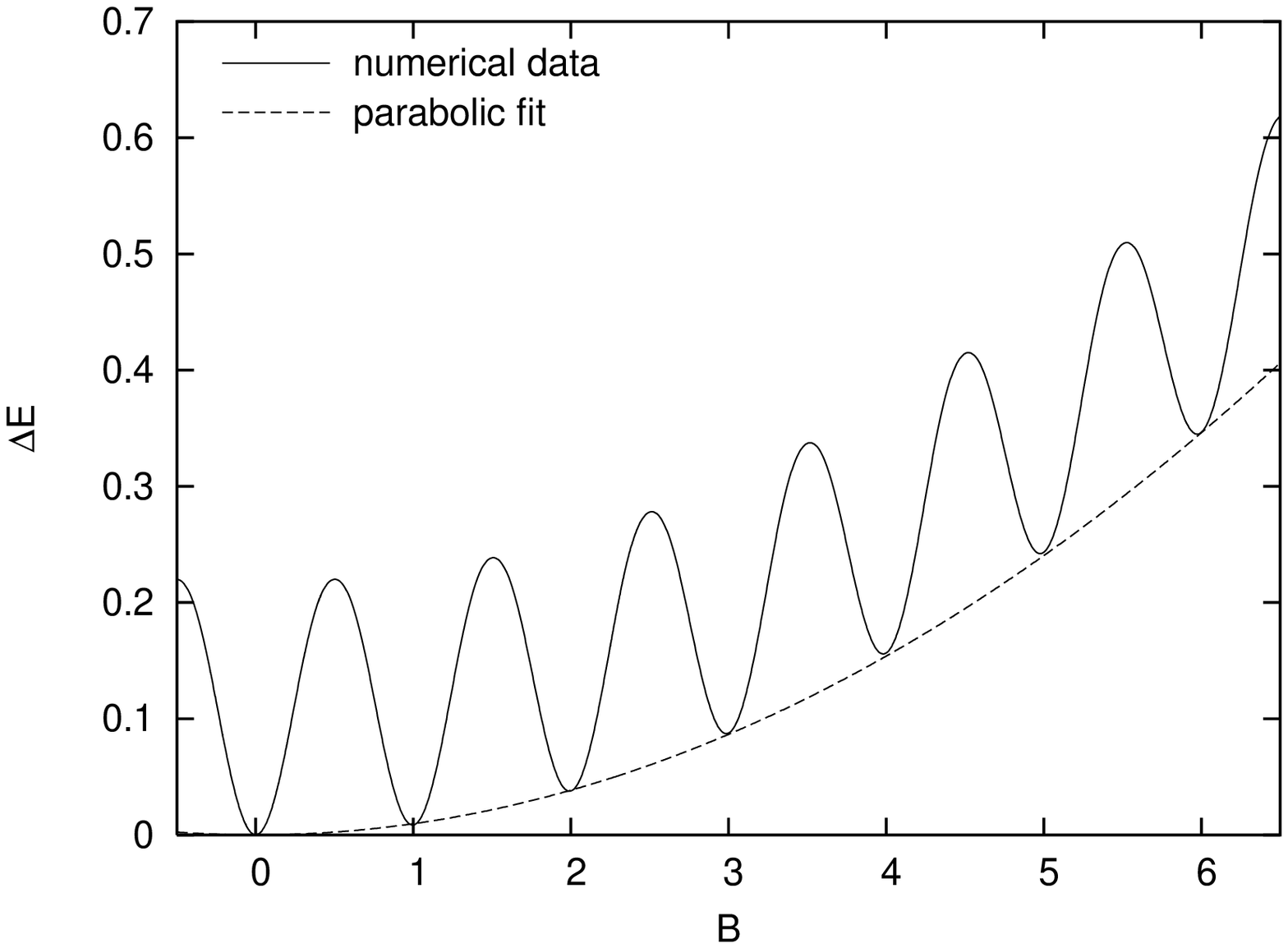}
\caption{Static energy in units of normal ``bare'' Skyrmion mass
$\msk$ as a function of the baryon number. Here $F_\pi^2/8v^2=0.0001$,
$\zeta^2/v^2=0.2$; the energy is referenced from the monopole mass:
$\Delta E=(E-\mm)/\msk$.}
\label{added}
\end{figure}

\begin{figure}[htb]
\centering \includegraphics[height=3.in]{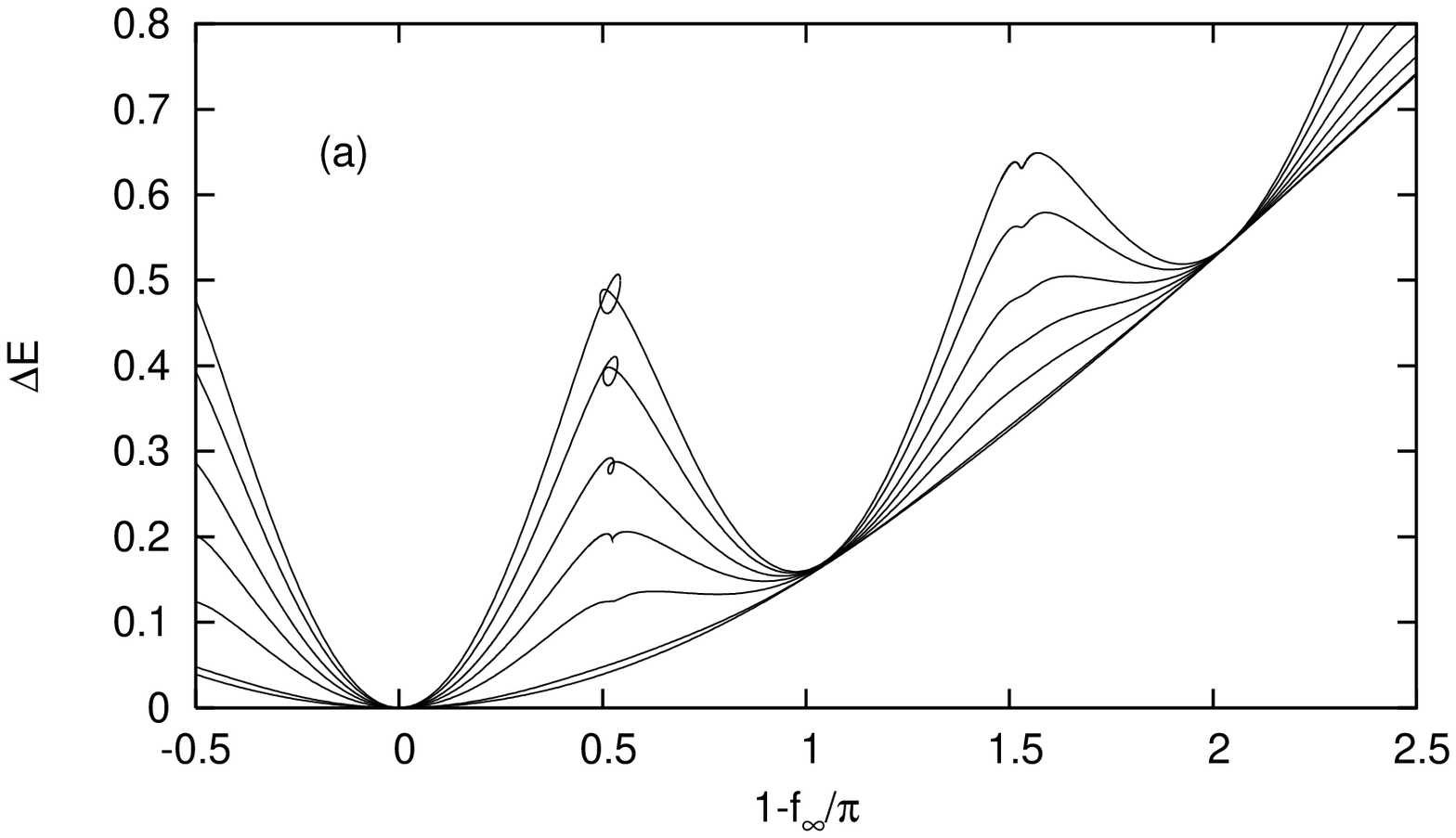}
\centering \includegraphics[height=3.in]{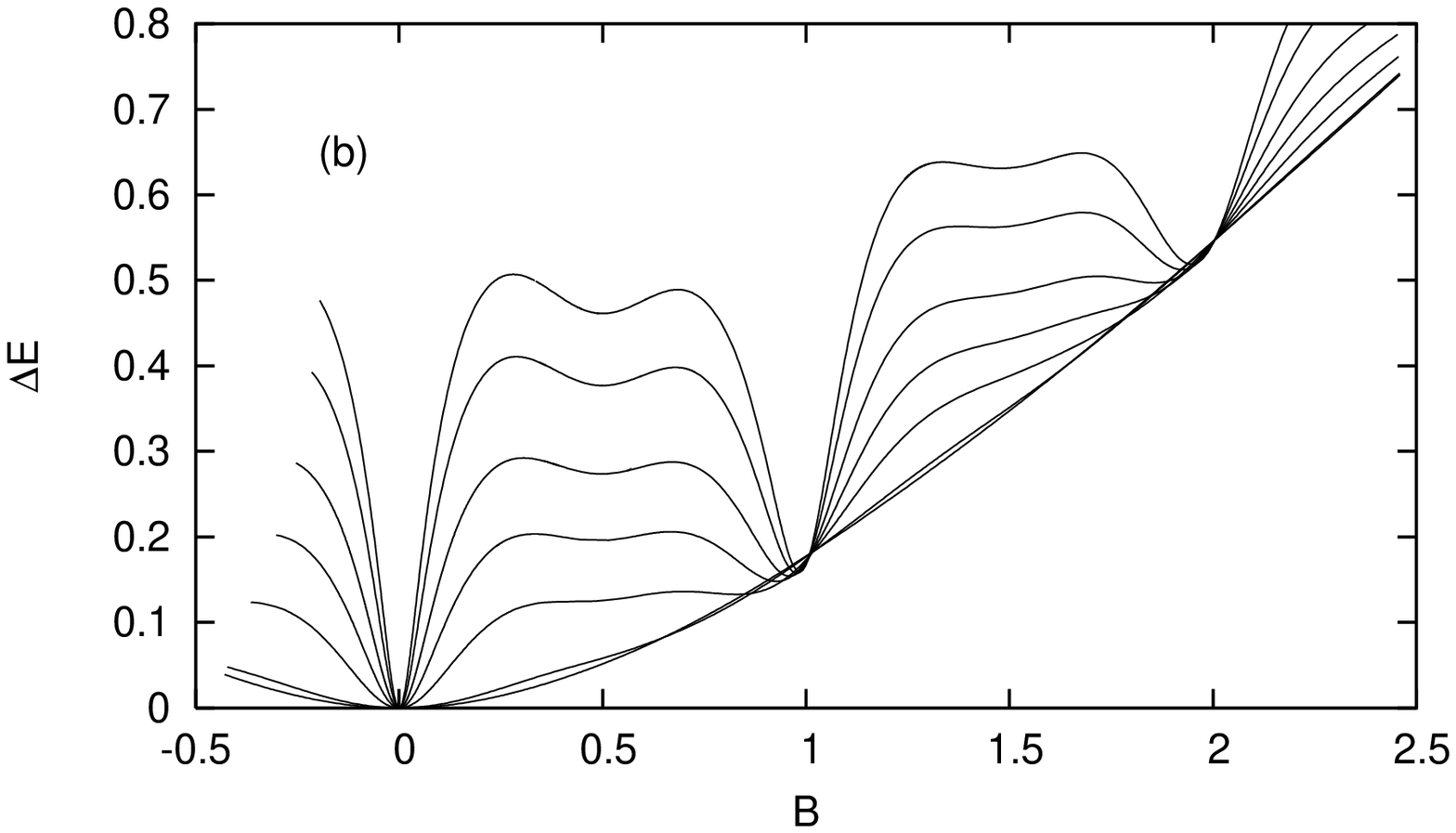}
\caption{If all mass scales are of the same order, the model becomes
  strongly coupled, which is demonstrated by the bifurcations in the energy
  dependence on $\fax$, plot (a). However bifurcations do not
  appear in energy versus the full baryon number B,
  plot (b). Curves plotted are for $\kappa_1=1$, $\kappa_2=0.1$,
  $(\zeta/v)^2=$ (top to bottom): 100, 70, 40, 22.36 (bifurcation
  point), 10, 1, 0.1 .}
\label{fig2}
\end{figure}

\begin{figure}[htb]
\centering \includegraphics[height=3.5in]{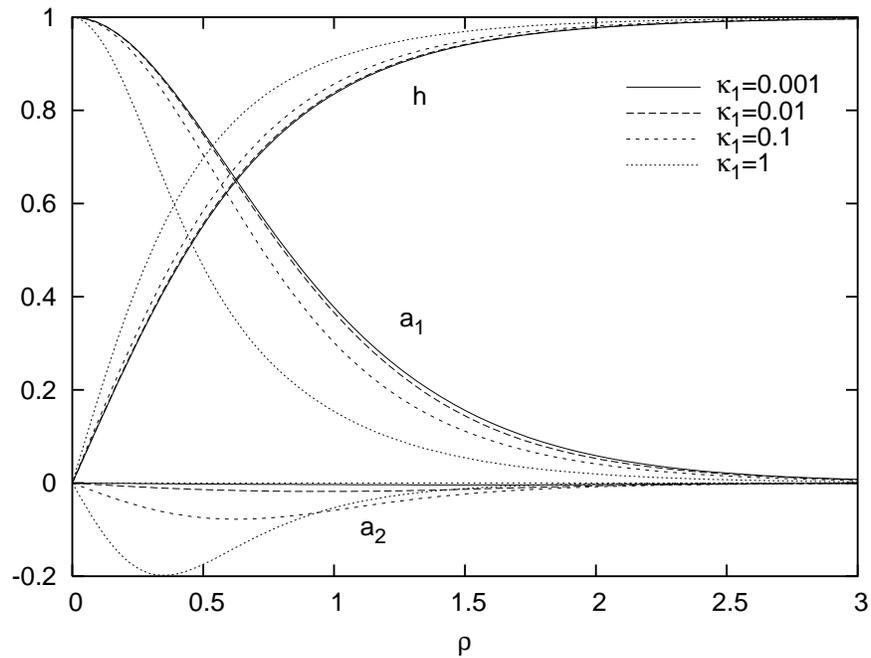}
\caption{At $\zeta/v, \kappa_1/\kappa_2 \ll 1$ the profile functions
$a_1$, $a_2$ and $h$ reach the free monopole limit and are not affected
by the low-energy fields $U$ and $\Psi$. We call this the spectator
monopole regime. For all curves $\kappa_2=0.1$, $\fax=0$ and
$(\zeta/v)^2=10^{-7}$. Normalization errors of Fig.~5, Ref.~\1 are
corrected.}
\label{fig1}
\end{figure}

\begin{figure}[htb]
\centering \includegraphics[height=3.5in]{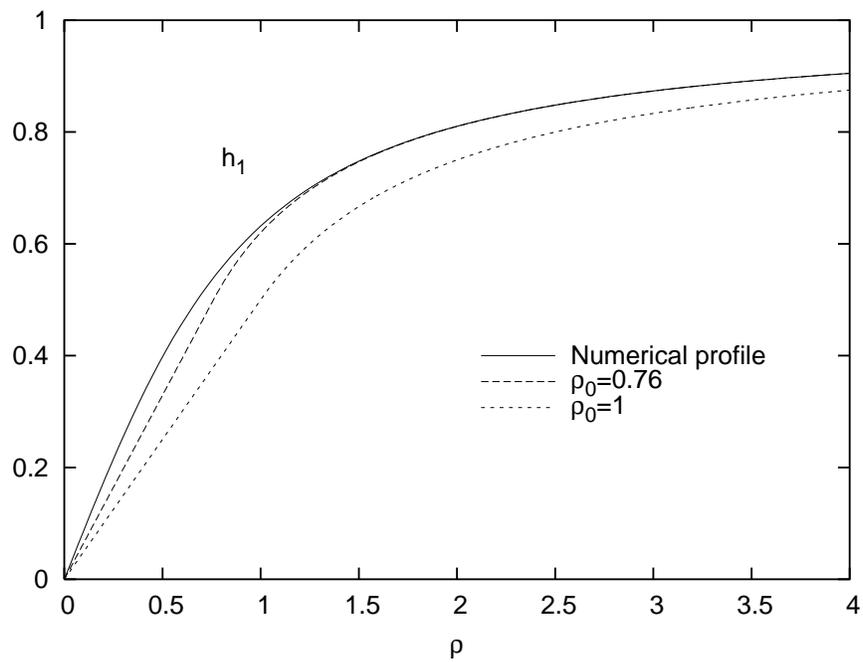}
\caption{The profile of the ``sphaleron'' -- a topologically-nontrivial
  configuration of the $\Psi$-field as compared to the analytic estimate,
Eq.~(\ref{sph}), at different values of $\rho_0$. The profile function
$h_2$ stays below $2\cdot 10^{-4}$.}  
\label{fig4}
\end{figure}

\begin{figure}[htb]
\centering \includegraphics[height=3.5in]{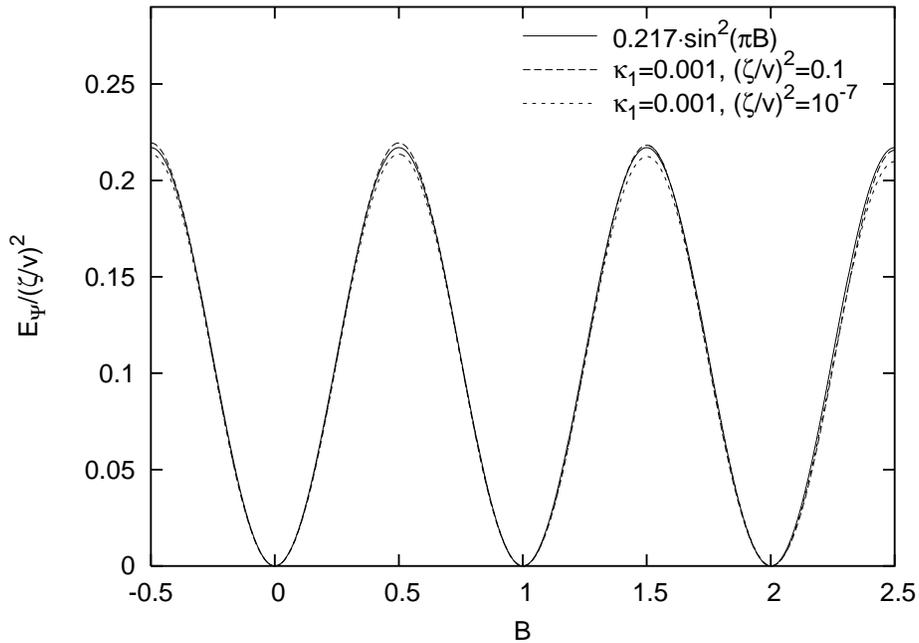}
\caption{After normalization by prefactor $(\zeta/v)^2$, the
contribution of the $\Psi$-field to the full energy (\ref{energyrad1})
is in perfect agreement with the $\sin^2\pi B$ curve,
Eq.~(\ref{periodic1}). However, the normalized sphaleron mass
(\ref{esph}) is estimated to be $\rho_0/2\sim 0.38$ instead of the actual
value 0.217 . The curves shown represent maximal deviations from
Eq.~(\ref{periodic1}) over all numerical data available.  In the physical
limit, this new periodic structure dominates the total energy.}
\label{fig5}
\end{figure}

\begin{figure}[htb]
\centering \includegraphics[height=3.5in]{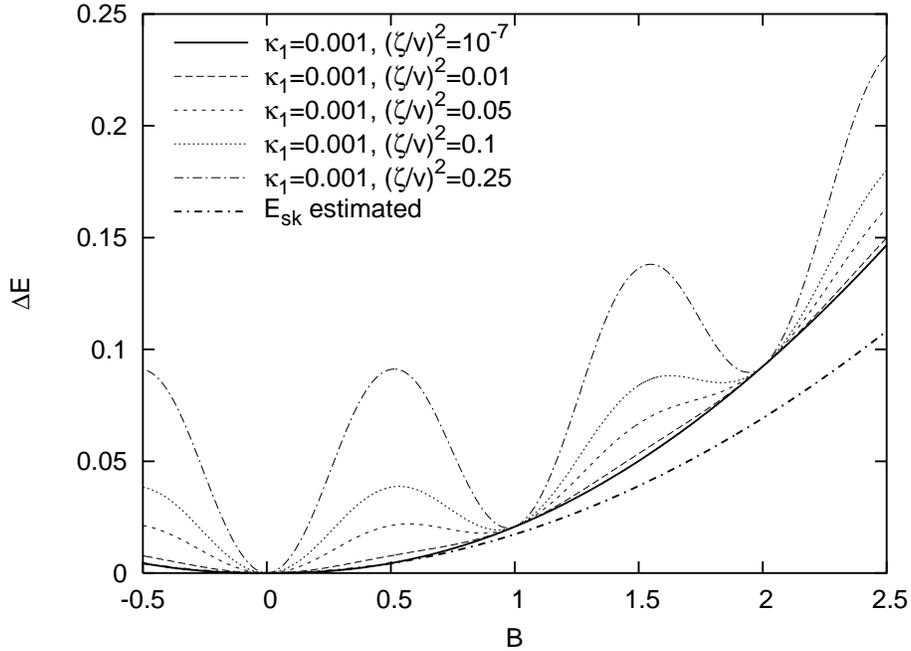}
\caption{In the spectator approximation the energy $\Delta E$, Eq.~(\ref{deltae}),
is equal to ``thin'' Skyrmion energy $\esk$, solid line (identical to
Fig.~1 of \1), plus the periodic contribution of the $\Psi$-field,
Fig.~\ref{fig5}. Once the latter is of order of $\zeta^2/v$
while $\esk\sim F_\pi^2/v$, see Eqs.~(\ref{eskyrm}) and (\ref{esph}),
in the physical limit the barriers are much higher than the Skyrmion
energy, which prevents Skyrmion decay along the static path. The lower 
line is the estimate of $\esk$, Eq.~(\ref{eskyrm}).}
\label{fig6}
\end{figure}

\begin{figure}[htb]
\centering \includegraphics[height=3.5in]{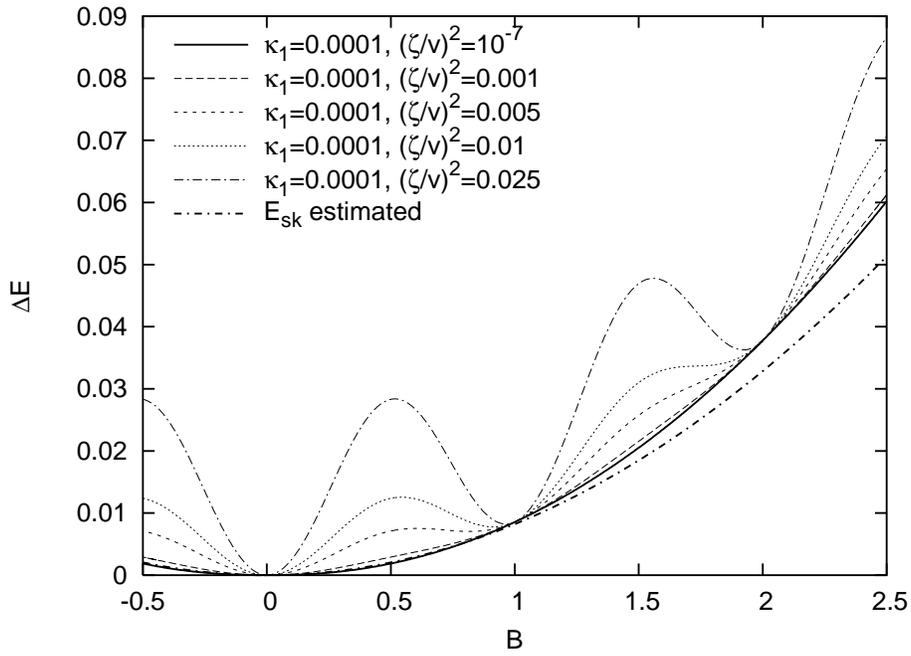}
\caption{Same as Fig.~\ref{fig6} for smaller $\kappa_1=8F_\pi^2/v^2$
and identical ratios of $\zeta^2/F_\pi^2$. Note that the relative
height of the barriers remains similar to Fig.~\ref{fig6} and
the estimate for $\esk$ gets closer to the numerical curve (solid line),
what demonstrates the onset of the spectator limit.}
\label{fig7}
\end{figure}

The existence of the barrier between local minima of static energy
 does not mean at all that the Skyrmion decay cannot take
place in the Skyrmion-monopole scattering
in the model at hand. Here the incoming ``bare'' Skyr\-mion becomes
unstable in the presence of the monopole and thus does not correspond
to any static solution. The ``bare'' Skyrmion is much heavier 
than its ``thin'' static counterpart with unit
baryon charge, the latter 
concentrating in the vicinity of the monopole core. 
The actual decay of the Skyrmion is a dynamical
process, not a quasistationary one, and the corresponding
 path goes much higher
than the $E(B)$ curve of Fig.~\ref{added}.  In other words,
to get into one of the stable local minima at integer $B$, the Skyrmion
must release almost all of its energy while preserving the baryon
number, which in the monopole background is no longer stabilized by
boundary conditions at infinity. This appears to be highly unlikely.

\bigskip

\noindent
{\bf Acknowledgments} The authors are grateful to V.B.~Kopeliovich,
A.V.~Smil\-ga, J.J.M.~Verbaarschot and A.R.~Zhitnitsky for stimulating
discussions. This work was carried out in the framework of projects
SC/00/020 and IC/01/073 of Enterprise--Ireland. The work of V.R. is
supported in part by Russian Foundation for Basic Research grant
02-02-17398 and U.S. Civilian Research and Development Foundation
(CRDF) award RP1-2364.


\end{document}